\documentclass[twocolumn, aps, prb, longbibliography]{revtex4-1}
\usepackage[normalem]{ulem}

 \pdfoutput=1

\usepackage{graphicx}
\usepackage{xcolor}
\usepackage{epsfig}
\usepackage{subfigure}
\usepackage{natbib}
\usepackage{array}
\usepackage{amsmath}
\usepackage{enumitem} 

\makeatletter

\begin{document}
\newcolumntype{M}{>{$}c<{$}}

\title{Defect-induced $4p$-magnetism in layered platinum diselenide}
\author{Priyanka Manchanda}
\author{Pankaj Kumar}
\author{Pratibha Dev}
\affiliation{Department of Physics and Astronomy, Howard University, Washington, D.C. 20059, USA}

\begin{abstract}


Platinum diselenide (PtSe$_{2}$) is a recently-discovered extrinsic magnet, with its magnetism attributed to the presence of Pt-vacancies.  The host material to these defects itself displays interesting structural and electronic properties, some of which stem from an unusually strong interaction between its layers. To date, it is not clear how the unique intrinsic properties of PtSe$_2$ will affect its induced magnetism. In this theoretical work, we show that the defect-induced magnetism in PtSe$_{2}$ thin films is highly sensitive to: (i) defect density, (ii) strain, (iii) layer-thickness, and (iv) substrate choice. These different factors dramatically modify all magnetic properties, including the magnitude of local moments, strength of the coupling, and even nature of the coupling between the moments. We further show that the strong inter-layer interactions are key to understanding these effects.  A better understanding of the various influences on magnetism can enable controllable tuning of the magnetic properties in Pt-based dichalcogenides, which can be used to design novel devices for magnetoelectric and magneto-optic applications. 

\end{abstract}

\maketitle

\section{Introduction}
The recent discovery of magnetism in two dimensional (2D) layered materials~\cite{Gong2017, Huang2017, Deng2018,Ohara2018} has sparked renewed interest in one of the oldest and most well-studied emergent phenomenon within solid state physics: collective magnetism. Not only do these 2D magnets circumvent the Mermin-Wagner theorem, which forbids long-ranged order in lower dimensions through magnetic anisotropy, but they also possess an interesting tunability of their magnetic properties that is facilitated by their 2D nature.  
For instance, magnetic ordering in CrI$_{3}$ thin films~\cite{Huang2017, Thiel2019} has been shown to depend on the number of layers. It was also shown that the nature of magnetic coupling in a CrI$_{3}$ bilayer~\cite{Huang2018} can be tuned by electric fields (gating).  Theoretical work on  CrX$_{3}$ (X = Cl, Br, I) monolayers~\cite{Webster2018} also indicates that strain can be used to tune magnetic properties. Hence, 2D magnets and their tunable properties offer opportunities to study the phenomena of local moment formation and collective magnetism in low-dimensions, as well as to design devices for spintronics, magneto-optics, quantum-information and -sensing applications. These 2D magnets can be: (i) intrinsic in nature, such as CrI$_{3}$, or (ii) extrinsic in nature. In the latter category, we include all 2D materials, otherwise non-magnetic, in which magnetism can be induced through: (i) defect engineering (vacancies, substitutionals, creating edges/nanoribbons)~\cite{Yazyev2007,Liu2007,Magda2014,Dev2015}, (ii) intercalation between layers by magnetic species~\cite{Dresselhaus1986,Morosan2007,Bointon2014,Kumar2017,LiKerui2019}, or (iii) proximity effects (2D crystal placed on a magnetic substrate)~\cite{Averyanov2018, Karpiak2019}. With only a few known intrinsic 2D magnets, there is a strong motivation to use the  aforementioned strategies to induce collective magnetism in non-magnetic 2D crystals.


Amongst extrinsic 2D magnets, long-ranged magnetism was recently discovered in platinum diselenide (PtSe$_{2}$) thin films~\cite{Avsar2019}. In this joint experimental-theoretical work, Avsar \textit{et al.} attributed the observed magnetism to the presence of platinum vacancies (V$_{\mathrm{Pt}}$)~\cite{Avsar2019}, which had also been previously investigated using density functional theory (DFT)~\cite{Gao2017}. 
In addition, several theoretical works have proposed other strategies for inducing a magnetic moment in PtSe$_{2}$ monolayers, such as: (i) combining strain with hole doping to induce magnetism~\cite{Zulfiqar2016}, (ii) Se-vacancy in the simultaneous presence of strain~\cite{Zhang2016}, (iii) hydrogenation on Se sites~\cite{Manchanda2016}, and (iv) doping with transition metal elements~\cite{Kar2019}. Most of these theoretical works used freestanding monolayers of PtSe$_{2}$ placed in vacuum, ignoring the effects of layer-thickness, or the presence of a substrate. However, experiments indicate that the magnetic properties in 2D crystals are strongly affected by layer thickness and substrates~\cite{Bonilla2018,Avsar2019, Avsar2020}. For example, in a more recent paper by Avsar \textit{et al.}~\cite{Avsar2020}, the authors discovered that magnetic ordering in PtSe$_{2}$ is antiferromagnetic in a monolayer, while it is ferromagnetic in a bilayer. This is an important result, showing layer-thickness dependence of V$_{\mathrm{Pt}}$-induced magnetism in PtSe$_{2}$. The importance of taking layer-thickness and/or substrates into account in studies of 2D materials, in general, was also recently highlighted in other careful theoretical studies of different 2D crystals~\cite{Dev2014, Manchanda2020}.

 In this work, out of the different proposed ways of inducing magnetism in PtSe$_2$ thin films, we chose to use platinum vacancies, as these defects have been experimentally implicated in the observed magnetic ordering in PtSe$_{2}$~\cite{Avsar2019,Avsar2020,Ge2020}.  
 We explored the dependence of the defect-induced magnetism on several factors: (i) vacancy concentration, (ii) strain (iii) layer-thickness, and (iv) substrate choice.
Using DFT-based calculations, we find that the magnitude of the magnetic moment and the exchange coupling between the moments are modified with vacancy concentration as well as strain. Moreover, the magnetic properties depend significantly on layer thickness/substrate due to the strong interlayer interaction between the layers in the presence of Pt-vacancies. Our results show that accounting for the ``real-world conditions", such as layer thickness, strain and the presence of a substrate, is imperative when predicting the magnetic properties of 2D materials.

\begin{figure*}[ht]
	\centering
	\includegraphics[width=0.72\textwidth]{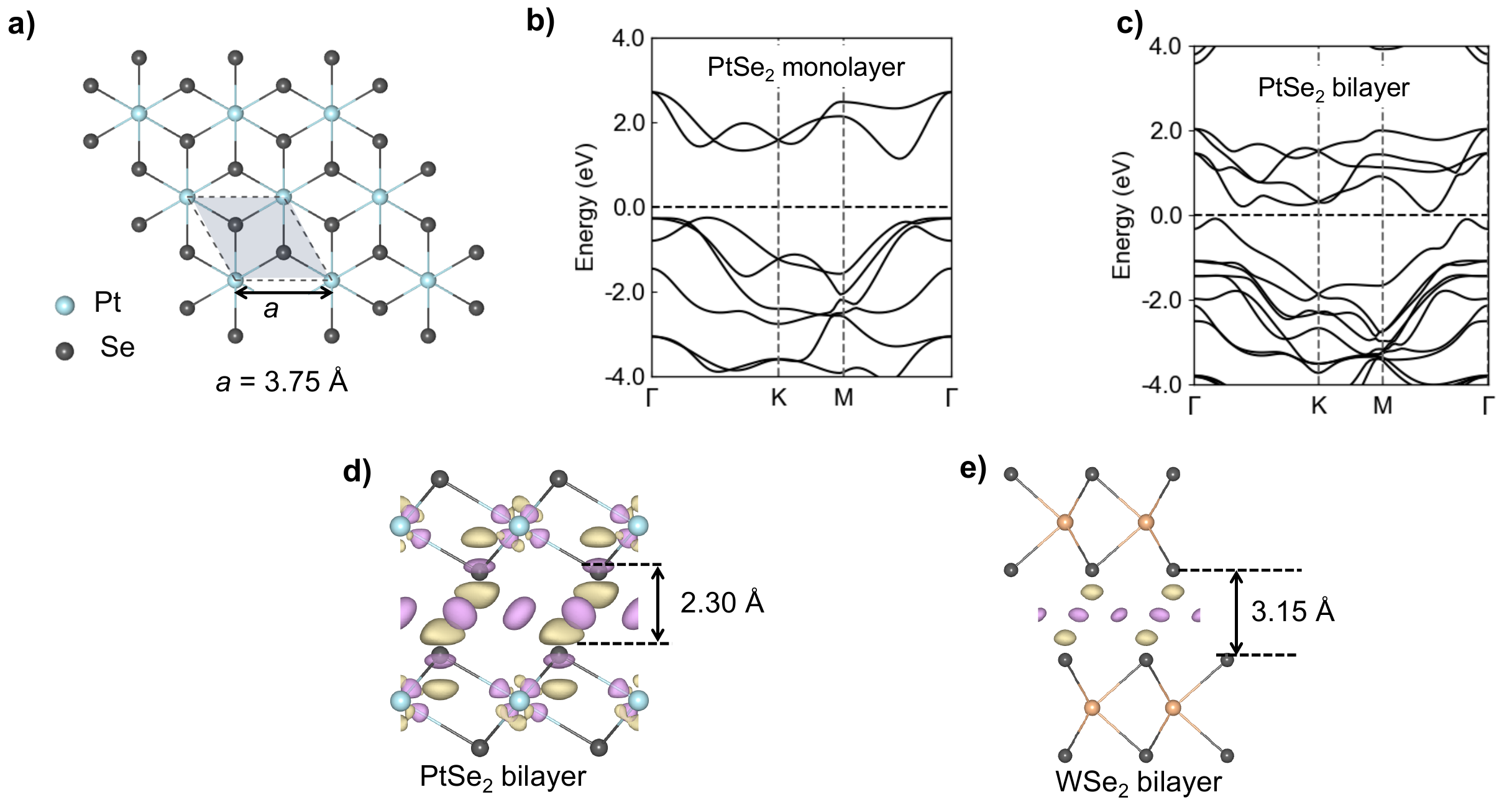}
	\caption{Pristine PtSe$_{2}$ properties. (a) Structure (top view) of pristine \textit{T}-phase PtSe$_{2}$ monolayer. The dashed area shows the primitive unit cell for PtSe$_{2}$. Band structures for a PtSe$_{2}$ (b) monolayer and (c) bilayer, showing layer dependence of electronic structure properties. The charge density difference plot, $\Delta\rho = \rho\textrm{(bilayer)}-\rho\textrm{(top-layer)}-\rho\textrm{(bottom-layer)}$, for (d) PtSe$_{2}$ and (e) WSe$_{2}$ bilayers, respectively, showing larger interlayer interaction in the former structure versus the latter structure. The pink and yellow isosurfaces represent charge accumulation and depletion, respectively.}
	\label{fig:figure1}
\end{figure*}

\section{Calculation Details}
Our spin-polarized DFT calculations were performed using the Vienna Ab-initio Simulation Package (VASP)~\cite{Kresse1996}. A small subset of the calculations were also performed using the Quantum Espresso Package~\cite{Giannozzi2009}. In all of the calculations, the generalized gradient approximation (GGA) of Perdew-Burke-Erzerhof (PBE)~\cite{Perdew1996} was used to account for exchange correlation effects.  To correctly describe the interlayer interactions in PtSe$_{2}$ thin films, Grimme's DFT-D3 van der Waals (vdW) corrections was used~\cite{Grimme2010}. The kinetic energy cut-off was set to 500\,eV. To study the effect of defect concentrations, we created V$_{\mathrm{Pt}}$ in $n \times n \times 1$ supercells, with  $n$=3, 4 and $5$, corresponding to the defect concentrations of 11.11\%, 6.25\% and 4.00\%. For the 11.11\%, and 6.25\% vacancy concentrations, we varied the layer thickness (up to 4-layers), placing the vacancy in the topmost layer. The \textit{k}-point sets used in all of our calculations were equivalent to at least a $21\times21\times1$-grid for the unit cell of PtSe$_{2}$. To study the effects of substrates, we considered three representative substrates: graphene [used in the experiment by Avsar \textit{et al.}~\cite{Avsar2020}], hexagonal boron nitride [hBN], and copper [Cu(111)]. For the PtSe$_{2}$/hBN and PtSe$_{2}$/graphene heterostructures, we used a $4 \times 4 \times 1$ supercell PtSe$_{2}$ placed on a $6 \times 6 \times 1$ supercell of hBN or graphene (lattice mismatch: 0.33\% and $-1.60$\%, respectively). For PtSe$_{2}$/Cu(111), we used $4 \times 4 \times 1$ supercell of PtSe$_{2}$ on a 4-atom thick layer of Cu(111), with a lattice mismatch of 2.62\%. For all the structures, energy convergence criterion was set to be better than 10$^{-6}$ eV and the atomic relaxations were carried out until the Hellmann-Feynman 
forces were smaller then 10$^{-2}$ eV/\AA{}. The layers were separated by 25 \AA{} of vacuum to eliminate spurious interactions between the periodic images. 


\section{Results and Discussion}




 Figure~\ref{fig:figure1}(a) shows the optimized structure of a pristine PtSe$_{2}$ monolayer, for which the lowest energy phase is the octahedrally-coordinated \textit{T}-phase. Before investigating defect-induced magnetism in 1T-PtSe$_{2}$, it is instructive to consider the unique properties of the pristine structure, which set PtSe$_{2}$ apart from more extensively-investigated group-VI TMDs. Recent experimental~\cite{Ciarrocchi2018} and theoretical~\cite{Villaos2019,Zhang2017} studies showed a dramatic change in its electronic structure properties as a function of thickness. This material, which prefers $AA$-stacking, is metallic in bulk, semi-metallic as a trilayer, and semiconducting as a bilayer or a monolayer. Figures~\ref{fig:figure1}(b) and (c) are plots of the band structures of a pristine monolayer and a bilayer, respectively, showing the layer-dependence of the indirect band gap, which changes from about $1.40$\,eV for the former structure to about $0.20$\,eV in the latter structure. This strong dependence of electronic structure properties on the sample thickness stems from a strong interlayer interaction, resulting in much smaller distances between the layers. For example, in the case of a PtSe$_{2}$ bilayer, the binding energy, $E_{b}$, is $-0.199$\,eV/f.u. [with $E_{b}=E_{bilayer} -E_{top-layer} - E_{bottom-layer}$, $E_{X}$ = total energy of system $X$, and formula unit abbreviated as f.u.], and the calculated interlayer distance is about $2.30$\,\AA{}. For comparison, the calculated binding energy for a WSe$_{2}$ bilayer is  $-0.148$\,eV/f.u., and the interlayer distance is about $3.15$\,\AA{}, where we have chosen WSe$_{2}$ as a representative material from  group-VI TMDs. The strong interlayer interaction in PtSe$_2$ can also be seen from the charge density difference ($\Delta\rho$) plot for the PtSe$_{2}$ bilayer shown in Fig.~\ref{fig:figure1}(d). Here, $\Delta\rho= \rho\textrm{(bilayer)}-\rho\textrm{(top-layer)}-\rho\textrm{(bottom-layer)}$, is a measure of the extent to which there is a charge redistribution upon formation of a bilayer, and hence, a measure of the interlayer interactions.  In contrast, for the case of WSe$_{2}$-bilayer shown in Fig.~\ref{fig:figure1}(e), the interaction is mostly van der Waals in nature and the structure shows minimal charge-redistribution.  In the subsequent discussion, we will show how the greater strength of interlayer interactions profoundly affect defect-induced magnetism in PtSe$_2$ layers.

 \begin{figure*}[ht]
	\centering
	\includegraphics[width=0.85\textwidth]{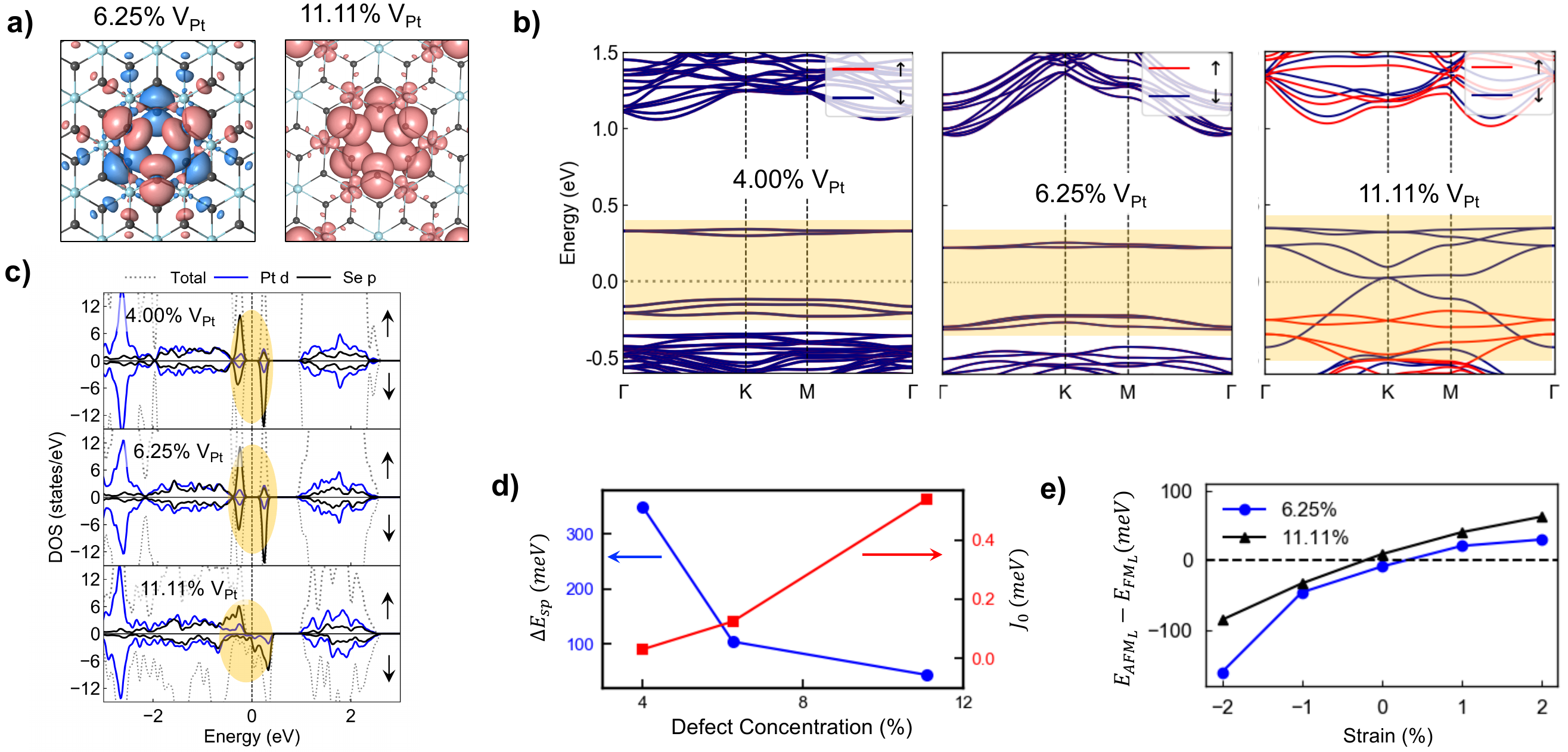}
	\caption{Properties of platinum vacancies in a PtSe$_2$ monolayer. (a) The isosurface plot of spin density distribution for V$_{\mathrm{Pt}}$= 6.25\% and 11.11\%. (b) Band structure for different defect concentrations, showing overall increase in the bandwidths (yellow highlight) with increasing defect concentration. (c) Density of states (DOS) for defective structures, showing the defect states (yellow highlight) around the Fermi level (used as the reference energy).  (d) Spin polarization energy per defect (left) and exchange coupling constant (right) as a function of defect concentration. Spin polarization energy shows a decrease in spin polarization as the overlap between the spin-split defect-states increases with increasing defect concentrations. (e) The computed 
$\Delta E_{L} = E_{AFM_{L}}-E_{FM_{L}}$ as a function of strain at defect concentrations of 6.25\% and 11.11\% for PtSe$_{2}$ monolayers. }
	\label{fig:figure2}
\end{figure*}

\subsection{Defect-induced magnetism in PtSe$_{2}$ monolayer}

 In PtSe$_{2}$, the platinum vacancy results in a unique reconstruction around the defect, in which the nearest-neighbor Se atoms move outwards away from the vacancy. The Se-Se distance increases both in-plane and out-of-plane as compared to those in a pristine PtSe$_{2}$ monolayer.  The extent of the outward movement depends on the defect concentration itself. In contrast to this outward movement in PtSe$_{2}$, in the case of a group-VI TMD, such as a WSe$_{2}$ monolayer, Se atoms surrounding the defect move closer to each other, as compared to a pristine WSe$_{2}$ monolayer. This inward movement of atoms leads to an increased overlap between the dangling bonds, resulting in a nonmagnetic structure. As a result, one does not associate metal vacancies in group-VI TMDs with induced magnetism, unlike those in PtSe$_2$.   

In a PtSe$_{2}$ monolayer, once we remove a platinum atom from the matrix, we find that the spin-polarized configuration is lower in energy than the non-magnetic configuration, with most of the magnetic moment originating from the partially occupied $4p$ orbitals of the six neighboring Se atoms. We also find that there are two possible spin-polarized configurations, corresponding to alignment of the local moments in the two Se planes. This is in agreement with the theoretical results reported by Avsar \textit{et al.}~\cite{Avsar2020}. The spins in the two Se-planes around the defects can be oriented parallel to each other, or they can be oriented antiparallel to each other. We refer to the former configuration as $FM_{L}$, and latter as $AFM_{L}$. Here, the subscript $L$ is used to indicate that we are referring to the distribution of the local moment in each defect center, and not to collective macroscopic magnetism in the system.  For a monolayer with 11.11\% defect concentration, the $FM_{L}$ configuration is lower in energy than the $AFM_{L}$ configuration by about 9.10\,meV. 
For the lowest energy ($FM_{L}$) configuration, we find that a total magnetic moment of $4\,\mu_{B}$ is introduced in the freestanding PtSe$_{2}$ monolayer. This value of total magnetic moment per defect is expected as the formal oxidation state of platinum is $+4$ in 1T-PtSe$_{2}$. Most of the magnetic moment originates from the partially occupied $4p$ orbitals of the six neighboring Se atoms ($\sim0.58\,\mu_{B}$ per atom), as shown in Fig.~\ref{fig:figure2}(a). For the two smaller defect concentrations -- 6.25\% and 4.00\% -- $AFM_{L}$ is lower in energy than the $FM_{L}$ configuration by $8.64$\,meV and $21.81$\,meV, respectively. Spin density isosurfaces for the 6.25\% and 11.11\% defect concentrations are plotted in Fig.~\ref{fig:figure2}(a), showing their respective preferred $AFM_{L}$ and $FM_{L}$ spin alignments. The spin density distribution is defined as: $\Delta \rho$ ($= \rho^{\uparrow} - \rho^{\downarrow}$), where $\rho^{\uparrow}$ and $\rho^{\uparrow}$ are charge densities in the two spin channels.

Figure~\ref{fig:figure2}(b) shows the band structures of defective PtSe$_2$ monolayers with three different defect concentrations. For the 4.00\% and 6.25\% concentrations, spin-up (majority) and spin-down (minority) states overlap due the antiferromagnetic coupling between the Se atomic planes, whereas the band structure for the largest defect concentration (11.11\%) shows spin-splitting of the defect states into the spin-up (majority) and spin-down (minority) states.  For the largest defect concentration, the bandwidths of defect states [highlighted in Fig.~\ref{fig:figure2}(b)] are larger due to the high defect concentration, which results in an increased interaction between the defects. This large dispersion can be seen in the widths of the highlighted defect-induced states around the Fermi level in the density of states (DOS) plot for the largest defect concentration in Fig.~\ref{fig:figure2}(c). As a result of the large dispersion, there is a considerable overlap between the two spin channels, resulting in a smaller spin-polarization energy (per defect) of $44.27$\,meV [see Fig.~\ref{fig:figure2}(d)].  Here, the spin polarization energy is defined as a difference in total energies of the non-magnetic ($E_{NM}$) and the most stable spin polarized configuration  ($E_{AFM_{L}/FM_{L}}$) structures: $\Delta E_{sp}=E_{NM}-E_{AFM_{L}/FM_{L}}$. For the smaller defect concentrations, the defect states are nearly dispersionless, as can be seen in the band structures and DOS plots in Figs~\ref{fig:figure2}(b) and (c), respectively. In turn, this results in larger spin polarization energies. The calculated values of spin-polarization energies (per defect) for the 6.25\% and 4.00\% concentrations are 103.80\,meV and 349.46\,meV respectively [see Fig.~\ref{fig:figure2}(d)]. These are large energies (compared to room temperature ($k_{B}T\approx25$\,meV), implying that the local moments will survive at room temperature. 

Formation of defect-induced local moments does not necessarily results in collective magnetism, with the latter being a result of the exchange interaction between the moments on different defect centers. In order to determine the strength and the nature of magnetic exchange coupling between the defect-induced moments, we determined the energy difference between the ferromagnetic (FM) and antiferromagnetic (AFM) alignments of the moments. In order to do so, we doubled the size of the supercells in one of the lateral directions, and depending upon how we initialized the moments, we obtained either the AFM or the FM alignment of moments on the neighboring defect centers. The difference in total energies for the two stable magnetic structures, $\Delta E= E_{AFM}-E_{FM}$, was then mapped onto an effective Heisenberg spin model, which gives the distance-dependent values of the exchange coupling constant. According to the Heisenberg model, taking into account only nearest-neighbor interaction: $\textit{H} = -J_{0}\sum_{\langle ij \rangle} S_i\cdot S_j$, where $J_0$ is the nearest-neighbor exchange coupling constant and $S_i$ is the spin induced by the $i^{th}$ platinum vacancy. This yields the expression: $J_{0}=\Delta E/8S^{2}$. The strength of the interaction between moments itself depends on the distance between the interacting moments and hence, on the distance between the defects.  The vacancy concentrations -- 4.00\%, 6.25\%, and 11.11\% -- considered here, correspond to defect-defect distances of 18.75\,{\AA}, 15.00\,{\AA}, and 11.25\,{\AA}, respectively. Figure~\ref{fig:figure2}(d)(right) shows that exchange coupling parameter is positive for all three distances, although it becomes smaller with distance. Even for defect-defect distance of 15.00\,{\AA} (6.25\% V$_{Pt}$), the exchange coupling constant is non-negligible (0.12\,meV), which indicates that there exists long-range magnetic ordering in defective PtSe$_{2}$ monolayers. In addition, a positive value of $J_0$ means that for smaller defect concentrations, all atomic-moments from one Se atomic plane will be aligned in one direction, which will be antiparallel to those on the other Se atomic plane.

\begin{figure*}[ht]
	\centering
	\includegraphics[width=0.77\textwidth]{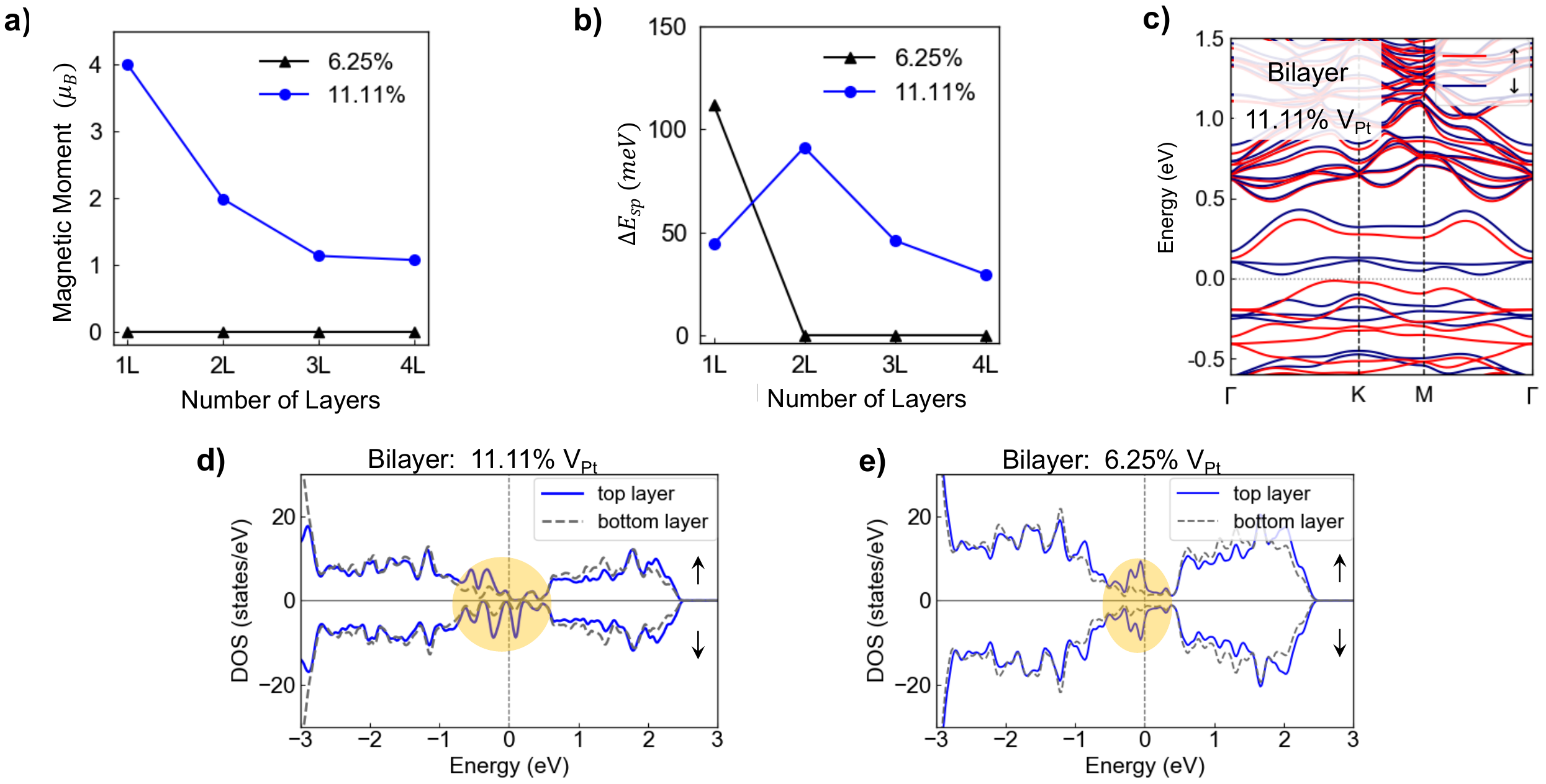}
	\caption{Magnetic properties of defective PtSe$_2$ thin films. Thickness-dependent (a) magnetic moment, and (b) spin-polarization energy per defect ($\Delta E_{sp}$) of PtSe$_{2}$ thin films with V$_{Pt}$ of 6.25\% and 11.11\%. (c) Band structure of the bilayer with 11.11\% defect concentration. (d) and (e) DOS projected on the top (defective) and bottom layers of a bilayer with 11.11\% and 6.25\% defect concentrations.}
	\label{fig:figure3}
\end{figure*}

As pointed out earlier, in a recent joint experimental-theoretical study~\cite{Avsar2020}, where PtSe$_2$-monolayer was placed on graphene, Avsar \textit{et al.} reported an experimentally measured antiferromagnetic ground state. They further estimated that their monolayer had a V$_{Pt}$  close to 6.25\%. At this defect concentration, we find that the macroscopic (collective) antiferromagnetic ground state corresponds to all moments on Se-atoms (surrounding every defect) on one face in an antiparallel direction relative to those on the other face of PtSe$_2$. This collective magnetic behavior corresponds to the antiparallel distribution of local moments on the two Se-planes ($AFM_{L}$) at each defect center. Hence, our results for the 6.25\% defect concentration agree with their experimental and theoretical results. However, their theoretical work predicted AFM$_L$ ordering to be the lowest energy configuration for all defect concentrations, in contrast to our results for the structure with 11.11\% defect concentration. A possible reason for this discrepancy can be due to the use of experimental lattice constants in their theoretical study, which may have introduced compressive strain in their structures. To further investigate if strain affects the ordering of local moments between the two Se-faces, we calculated $\Delta {E}_{L}=E_{AFM_{L}}-E_{FM_{L}}$ for monolayers with 6.25\% and 11.11\% defect concentrations by varying the strain from -2\% to +2\%.  The strain percentage is defined as $(a-a_{0})/a_{0}$, with $a_{0}$ and $a$ referring to the equilibrium and strained lattice constants, respectively. 
Fig. \ref{fig:figure2}(e) is a plot of $\Delta E_{L}$ as a function of strain. It shows that the magnetic ordering is very sensitive to applied strain, and even a small strain can switch the magnetic ordering. Overall, the ferromagnetic ordering becomes more favorable under applied tensile strain whereas antiferromagnetic ordering is preferred when compressive strain is applied, showing that any spurious strain in the calculation and/or experiment can affect the outcome. 

Long-ranged magnetism in 2D materials requires magnetic anisotropy in order to survive at finite temperature. In order to verify that collective magnetism will survive in PtSe$_2$, we calculated the magnetic anisotropy energy (MAE). MAE is defined as the difference in the total energies of the structures with magnetization that is parallel and perpendicular to the atomic plane: $MAE = E_\parallel -E_\perp$. Here, $E_\parallel $ and $E_\perp$ are obtained from non-collinear DFT calculations that take spin-orbit coupling (SOC) into account and employ fully-relativistic pseudopotentials. A positive (negative) value of MAE indicates a perpendicular (in-plane) easy axis. These calculations are computationally demanding and thus, were performed only for the two highest defect concentrations. The calculated MAE values are 13.73\,meV and 7.96\,meV for the defect concentrations of 6.25\% and 11.11\%, respectively. These results suggest that PtSe$_2$ monolayers should possess a magnetic ground state with out-of-plane magnetization.


\begin{figure*}[ht]
	\centering
	\includegraphics[width=0.77\textwidth]{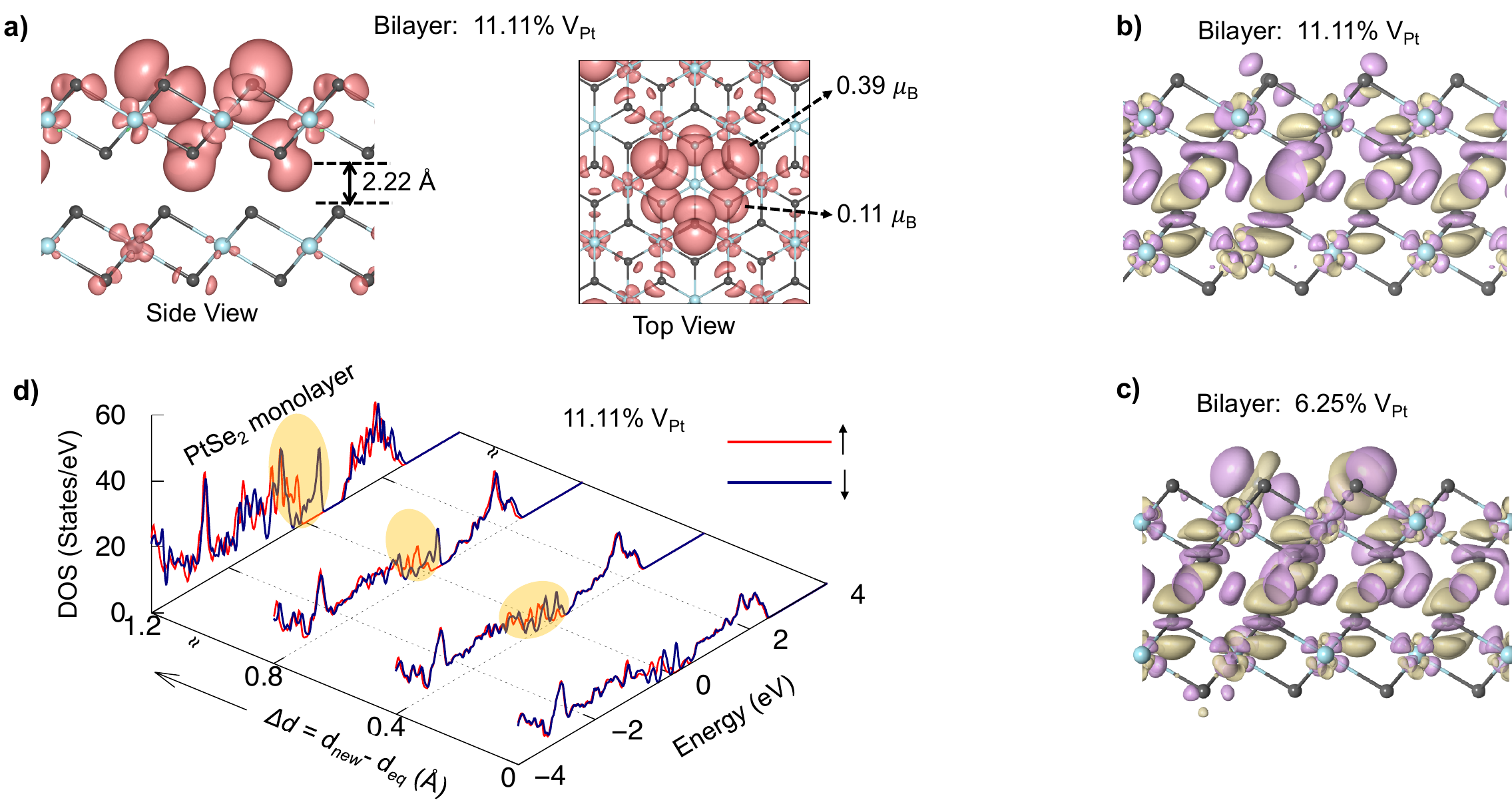}
	\caption{(a) Side and top views of the spin density distribution of a PtSe$_{2}$ bilayer with V$_{Pt}$ concentration of 11.11\%. (b, c) Charge density difference plot of PtSe$_{2}$ bilayer with V$_{Pt}$ concentrations of 11.11\% and 6.25\%, showing larger rearrangement of charges for two of the defect-concentrations. The pink and yellow isosurfaces represent charge accumulation and depletion, respectively. The isosurface value is taken as 0.001 e/\AA\textsuperscript{3}. (d) DOS projected onto a defective PtSe$_{2}$ layer within a bilayer with 6.25\% defect concentration as a function of the rigid shift, $\Delta d$, between the two layers. The rigid shift is given by: $\Delta d = d_{new}-d_{eq}$, where $d_{eq}$ is the distance between the two layers at equilibrium. Also shown is the DOS for the defective monolayer for comparison. The spin-split states induced by the defect are highlighted.}
	\label{fig:figure4}
\end{figure*}

\subsection{Thickness dependence of defect-induced magnetism in PtSe$_{2}$}

The defect-induced magnetic properties of PtSe$_{2}$ are expected to be affected by not only the vacancy concentration and strain, but also the layer thickness due to strong interlayer interactions. In this work, we considered up to 4 layer-thick PtSe$_{2}$ films. Due to the computational costs, we restricted this part of the study to only two V$_{Pt}$ concentrations: 6.25\% and 11.11\%. In each case, V$_{Pt}$ was created on the topmost layer only since our calculations indicated that a vacancy in one of the inner layers of a PtSe$_{2}$-stack does not yield a net magnetic moment due to strong interlayer interaction between the layers, and the system remains nonmagnetic. Fig.~\ref{fig:figure3}(a) shows the effect of PtSe$_{2}$ layer thickness on the magnetic moment for the considered systems. As discussed in the previous section, in the case of a monolayer, the value of the net magnetic moments are $4.00\,\mu_{B}$ and $0.00\,\mu_{B}$ for 11.11\% and 6.25\% defect concentrations, respectively [see Fig.~\ref{fig:figure3}(a)]. For the higher defect concentration of 11.11\%, the structure remains magnetic as we add more PtSe$_2$ layers, although the moment decreases with the number of layers. For example, the magnetic moment reduces from $4.00\,\mu_{B}$ in a monolayer to a value of $1.99\,\mu_{B}$ in a bilayer. On the other hand, in the case of 6.25\% defect concentration, the system becomes non-magnetic for all thicknesses beyond a single-layer [see Figure~\ref{fig:figure3}(a)].  

Figure~\ref{fig:figure3}(b) gives spin polarization energy as a function of layer-thickness for different concentrations of vacancies. For the smaller defect concentration of 6.25\%, the structure becomes non-magnetic as thickness is increased beyond a single layer. This is reflected in the spin polarization values that reduce to zero [see the black curve in Fig.~\ref{fig:figure3}(b)].  The behavior of spin polarization energy values for the higher defect concentration (11.11\%) is more interesting.  At this concentration, the calculated spin-polarization energy for a bilayer shows an unexpected increase as compared to that obtained for a freestanding monolayer. The band structure for the bilayer with 11.11\% defect concentration, plotted in Fig.~\ref{fig:figure3}(c), reveals the interrelated factors that explain the unexpected increase of the spin-polarization energy. It shows a large reduction in the bandwidth of the defect-induced states (within the band gap) as compared to those in the band structure for the freestanding monolayer [see Fig.~\ref{fig:figure2}(b)] with the same defect concentration. The resulting spin-splitting between the two spin channels is relatively larger, explaining the increase in spin polarization energy. The change in the bandwidth of the defect-states in the bilayer as compared to those in the free-standing monolayer [shown in Fig.~\ref{fig:figure2}(b)] can be attributed to the differences in structural distortions around the defect. In turn, this is due to the substrate friction from the bottom layer, which changes the extent of relaxation around the defect in the defective top layer within a bilayer. Figure~\ref{fig:figure3}(d) is a plot of DOS for bilayers with 11.11\% defect concentrations, projected onto the top (defective) and the bottom layers. As was also seen in the band structure for the system, the highlighted defect states [see Fig.~\ref{fig:figure3}(d)] around the Fermi level (used as the reference energy) are narrower as compared to those for the freestanding monolayer with the same defect concentration [Fig.~\ref{fig:figure2}(c)]. This again implies greater localization of the defect states in the bilayer as compared to the monolayer. In the case of smaller defect concentration (6.25\%), the DOS projected onto the top and bottom layers within a bilayer reveals a symmetric DOS in the two spin channels [Fig.~\ref{fig:figure3}(e)], and hence, the quenching of magnetism for lower defect concentrations.  This result is at variance with the bilayer results by Avsar \textit{et al.}~\cite{Avsar2020}, where they experimentally find that the PtSe$_{2}$ bilayer with 6.25\% defect concentration to be  ferromagnetic. Their theoretical calculations also show that the structure is spin polarized, with the local moments in the two Se atomic planes arranged in parallel configuration, and a net magnetic moment of 1.33\,$\mu_B$. A possible reason for this discrepancy between our results and their reported theoretical results is that Avsar \textit{et al.} did not include the vdW correction in their calculation, and fixed the interlayer distance to the experimental bilayer value. However, the presence of defects increases the interlayer interactions, reducing interlayer distance [compare Figs.~\ref{fig:figure4}(a) and \ref{fig:figure1}(d)]. We will discuss a possible reason for discrepancy between our results for a bilayer with 6.25\% defect concentration and the experimental results~\cite{Avsar2020} in the following section.

To further understand how the presence of additional layers affects the magnetism in the top layer, we considered the spin density distribution, $\Delta \rho$. Figure~\ref{fig:figure4}(a) is a plot of the calculated $\Delta \rho$ for PtSe$_{2}$ bilayers with 11.11\% defect concentration. It shows a reduction in magnetic moment contributed by each of the Se-atoms that surround the defect. The magnetic moment contributed by each of the three surrounding Se-atoms located at the interface of the bilayer is about $0.11\,\mu_{B}$, while that contributed by each of the surrounding Se atoms on the surface is $\sim0.39\,\mu_{B}$ [see Fig.~\ref{fig:figure4}(a)]. This is quite different from the spin density profile of the PtSe$_{2}$ monolayer [Fig.~\ref{fig:figure2}(a)], where all six Se-atoms surrounding the Pt vacancy contribute equally to the total moment.   This near-quenching of the magnetism for the larger defect concentration, and its complete quenching for the bilayers with smaller defect concentrations, can be attributed to the following different, yet inter-dependent, factors: (i) changes in the structural distortion around the defect within a bilayer due to substrate-friction (here the substrate is another PtSe$_{2}$ layer), (ii) charge transfer from the bottom layer, and (iii) interface states due to strong interaction between the two layers. To investigate each of these effects sequentially, we followed the steps taken in Dev \textit{et al.}~\cite{Dev2014} and conducted the following series of tests on structures with 11.11\% defect concentration:

 (i) We first investigated the effect of the changed structural distortion in the defective layer, with these changes arising due to the presence of a neighboring layer. For this calculation, we considered a freestanding, defective PtSe$_{2}$ monolayer with the ``frozen-in" atomic positions obtained from the defective top layer at equilibrium within a bilayer. The magnetic moment remains 4.00\,$\mu_{B}$ in this monolayer.  These results indicate that the changes in the structure around a vacancy within a composite (as compared to a freestanding relaxed structure) do not play a role in reducing the magnetic moment.

 (ii) Another factor that can contribute to the reduction in moment is the calculated \textit{n}-doping of the defective top layer by the bottom PtSe$_{2}$ layer. Using Bader charge analysis~\cite{Henkelman2006}, the calculated charge transfer to the defective layer from its neighboring layer is 0.19\,e/supercell for the defect concentration is 11.11\%. Again starting with the hypothetical monolayer described in the previous paragraph, we doped the structures with 0.19\,e. Our calculations show charge doping results in a small reduction in magnetic moment to a value of 3.81\,$\mu_{B}$. Therefore, we find that although \textit{n}-doping of the top layer reduces the magnetic moment, this reduction is not significant.  

 (iii)  The third factor that we investigated is the stronger interlayer interaction within a bilayer in the presence of a defect. Our calculations show that the binding energies for the bilayers with 11.11\% defect concentration is $-0.287$\,eV/f.u. This is significantly higher than the binding energy for a pristine bilayer ($-0.199$\,eV/f.u.). The importance and strength of interlayer interaction can be seen in the charge density difference plot [Fig~\ref{fig:figure4}(b)]. There is an enhancement of charge rearrangement upon forming a bilayer when one of the PtSe$_{2}$ layers contains a defect [\textit{cf.} Fig.~\ref{fig:figure1}(d) for defect-free bilayer]. In addition, our calculations also show that the binding energy of the 6.25\% defect concentration is also high ($-0.264$\,eV/f.u.), resulting in large charge rearrangement in the structure. The structure with the  6.25\% defect concentration can be seen to have gone through even a larger charge rearrangement [Fig.~\ref{fig:figure4}(c)], compared to the structure with the larger defect concentration [Fig.~\ref{fig:figure4}(b)]. This might explain the complete quenching of magnetism for the structure with 6.25\% defect concentration. 
 
 In order to further test the effect of strong interlayer interaction, we started with the fully relaxed structure for a PtSe$_{2}$ bilayer with 11.11\% defect concentration with equilibrium inter-layer distance, $d_{eq}$, and moved the two layers apart in the steps of 0.2\,\AA. The magnetic moment increases as a function of the rigid shift, $\Delta$\textit{d}, and the system recovers the magnetic moment calculated for a freestanding layer (4.00\,$\mu_{B}$) when $\Delta d = 1.0$\,\AA. This can be seen from the  Fig.~\ref{fig:figure4}(d), which shows the DOS projected onto the top defective layer (with 11.11\% V$_{Pt}$) at different values of $\Delta$\textit{d}. As $\Delta d$ increases, the spin-splitting reappears, and the defect states around the Fermi energy become narrower, approaching the case of a freestanding monolayer. These results indicate that the strong interlayer interaction in a bilayer is a major contributing factor responsible for the reduction in magnetic moment.

\subsection{Substrate Effects}
\begin{figure}[ht]
	\centering
	\includegraphics[width=0.5\textwidth]{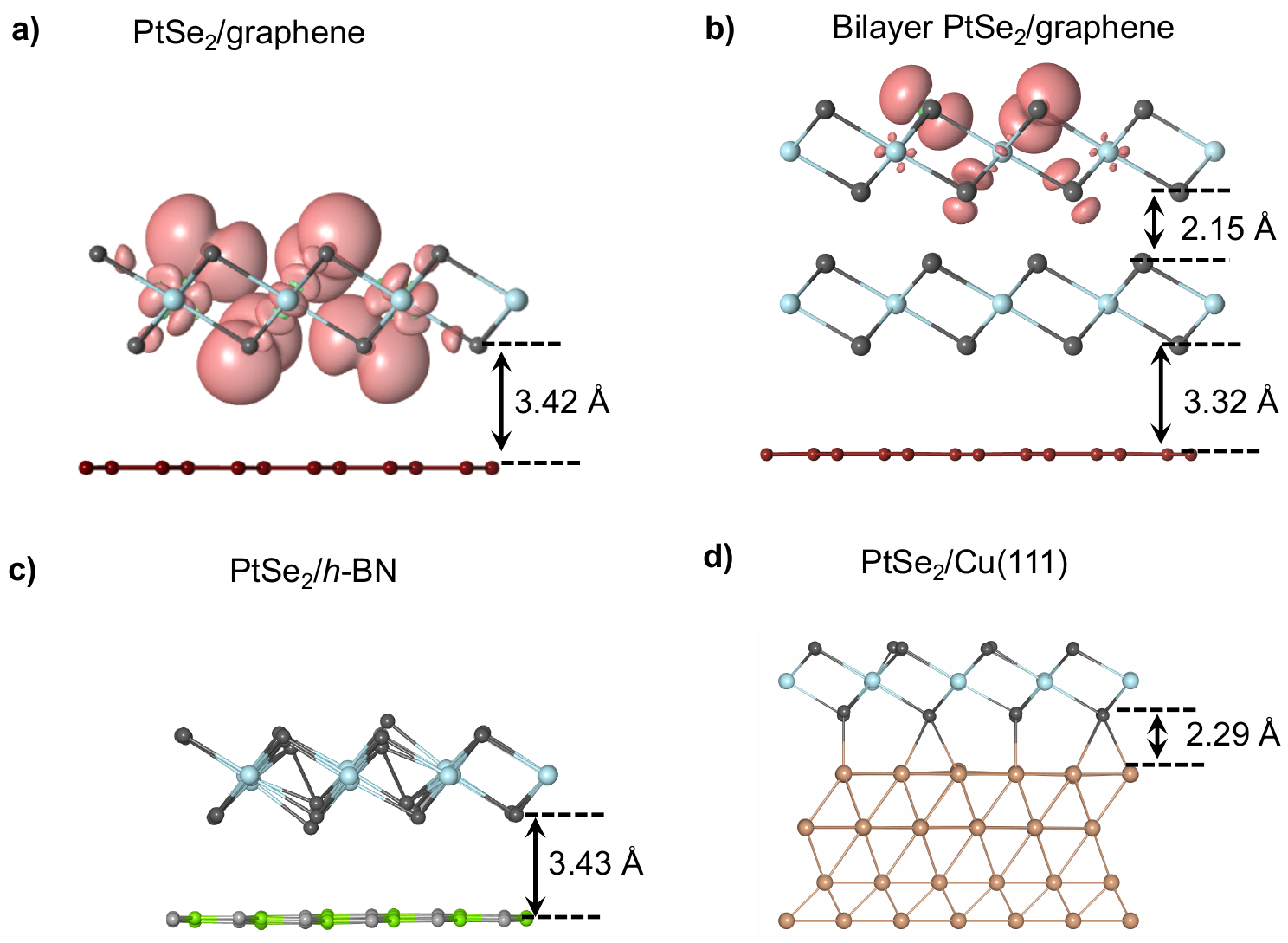}
	\caption{Effect of 2D and 3D substrates.  (a, b) Defective PtSe$_{2}$ monolayer and bilayer on graphene and their spin density distributions. (c) PtSe$_{2}$ on \textit{h}-BN. At equilibrium, the Se-atoms of the top and bottom layers surrounding the V$_{Pt}$ form bonds within the 2D PtSe$_{2}$/\textit{h}-BN heterostructure. (d) Within the mixed dimensional heterostructure consisting of PtSe$_{2}$ on a Cu(111) substrate, the strong interaction leads to the formation of covalent bonds.}
	\label{fig:figure5}
\end{figure}
Lastly, in most applications, the PtSe$_2$ layer(s) will be part of  2D or mixed-dimensional heterostructures. In fact, in the experimental report by Avsar \textit{et al.}~\cite{Avsar2020}, the authors used graphene as a substrate. Therefore, we investigated the effect of using conventional 2D or 3D substrates on the magnetic properties of defective PtSe$_{2}$. We chose three representative substrates: \textit{h}-BN and graphene, which are a 2D insulator and semimetal, respectively, and Cu(111), which is a 3D metal. The heterostructures are shown in Fig.~\ref{fig:figure5}.  We find that when a defective PtSe$_{2}$ monolayer (6.25\% defect concentration) is placed on graphene [see Fig.~\ref{fig:figure5}(a)], the system is spin-polarized, with $FM_{L}$ configuration lower in energy as compared to the $AFM_{L}$ configuration by 24.24\,meV. This is in contrast to the experimentally-observed antiferromagnetism reported by Avsar \textit{et al.}~\cite{Avsar2020} for the monolayer-PtSe$_{2}$/graphene heterostructure. Hence, our results agree with the experiment when we consider a freestanding defective PtSe$_{2}$ monolayer with 6.25\% defect concentration, but not when we take graphene into account. At this point, the exact reason for this difference with experiment remains unknown. However, we find that the lowest energy state for the PtSe$_{2}$/graphene heterostructure is also very sensitive to strain, and even a small compressive strain (-1\%) makes the system antiferromagnetic. Therefore, the presence of strain (e.g. due to other defects and/or substrate) might result in antiferromagnetic ground state, which was reported by Avsar \textit{et al.} for a PtSe$_{2}$/graphene heterostructure. To further mimic the experimental set-up, we also considered a structure where a defective PtSe$_{2}$ bilayer with 6.25\% defect concentration in the top layer is placed on graphene. In this case, no matter how the moments were initialized, we find that the structure is always ferromagnetic [see Fig.~\ref{fig:figure5}(b) for the spin density distribution] with a total magnetic moment of 0.71 $\mu_{B}$. Hence, in contrast to our results for the freestanding PtSe$_{2}$ bilayer, once we take the substrate into account, the structure is magnetic, in agreement with the experiment. We attribute the emergence of magnetism to competing interactions experienced by the buffer PtSe$_{2}$ layer sandwiched between the top layer and graphene.

In the cases of the \textit{h}-BN and Cu(111) substrates, the interactions of the substrates with the defective PtSe$_{2}$ layer result in the quenching of magnetism. We find that even though the interlayer distance between PtSe$_{2}$ and \textit{h}-BN is large ($\sim$\,3.43\,\AA) as compared to PtSe$_{2}$ bilayers, \textit{h}-BN strongly modifies the magnetic properties of PtSe$_{2}$ films. When PtSe$_{2}$  is placed on \textit{h}-BN, the reconstruction around the defect in the PtSe$_2$ layer results in bonding between Se atoms from the top and bottom faces around the vacancy [see Fig.~\ref{fig:figure5}(c)]. In order to investigate if the structure distortion in PtSe$_{2}$ on \textit{h}-BN results in non-magnetic structure, we studied the the distorted PtSe$_{2}$ with ``frozen-in" atomic positions obtained from the PtSe$_{2}$/\textit{h}-BN heterostructure. We find that such a structure is non-magnetic. 
When the PtSe$_{2}$ layer is placed on Cu(111), the monolayer interacts strongly with the copper substrate, which results in the formation of bonds between the two sub-systems in the composite structure [see Fig.~\ref{fig:figure5}(d)]. This strong interlayer interaction results in quenching of the magnetism in the monolayer. 

Hence, we find that substrates can profoundly affect the observed magnetic properties. In experiments, the outcome of using a substrate may change depending not only on the substrate-PtSe$_2$ interactions, but also a number of other factors, such as the defect concentration, local strain, and the number of PtSe$_2$ layers in the heterostructure.  Nevertheless, our proof-of-principle study for the 2D and mixed-dimensional heterostructures shows that a careful choice of substrates may be important to ensure survival of magnetism in the defective PtSe$_{2}$.

\section{Conclusions}
Using first-principles calculations, we studied magnetism in PtSe$_{2}$ layers that is induced by platinum vacancies. In a monolayer, the presence of V$_{Pt}$ results in a magnetic ground state, irrespective of the defect concentration. The dangling bonds around the defect are responsible for the local moment, with the largest contribution coming from the $4p$-orbitals of Se-atoms. These orbitals are an unusual source of unconventional magnetism, which is typically associated with unpaired electrons within the localized $2p$-orbitals of the second row elements~\cite{Dev2008,Dev2010,Dev2020}. On the other hand, the $4p$-orbitals give rise to highly dispersive bands and are not implicated in local moment formation in 3D solids. 
In PtSe$_2$, however, the $4p$-derived defect states surrounding the vacancy were found to be sufficiently localized to give rise to a magnetic structure. In fact, it was recently shown that the unpaired electrons in the $3p$- and $4p$-orbitals in 2D materials can indeed give rise to localized orbitals and hence, to a net magnetic moment due to the quantum-confinement effects in a 2D-structure~\cite{kumar2020}. We showed that this magnetism in defective PtSe$_2$ layers is strongly affected by the vacancy concentration, strain, thickness of the PtSe$_{2}$ layers, as well as the choice of the substrates. These different factors dramatically modify all magnetic properties, including the magnitude of local moments, distribution of local moments, strength of the coupling between the moments on different defect centers, and even the nature of the coupling between the moments.  Our results explain the dependence of the magnetic properties on different factors. Pt-based dichalcogenides are important candidate extrinsic magnets that can be used to design novel devices for magnetoelectric and magneto-optic applications. A better understanding of the various influences on magnetism will be important for controllably tuning their magnetic properties.

\section{Acknowledgements}
This work is supported by the National Science Foundation under Grant Number DMR-1752840. PD and PM acknowledge the computational support provided by the Extreme Science and Engineering Discovery Environment (XSEDE) under Project PHY180014, which is supported by National Science Foundation grant number ACI-1548562. For three-dimensional visualization of crystals and volumetric data, use of \textsc{VESTA 3} software is acknowledged.


\begin{thebibliography}{41}
\expandafter\ifx\csname natexlab\endcsname\relax\def\natexlab#1{#1}\fi
\expandafter\ifx\csname bibnamefont\endcsname\relax
  \def\bibnamefont#1{#1}\fi
\expandafter\ifx\csname bibfnamefont\endcsname\relax
  \def\bibfnamefont#1{#1}\fi
\expandafter\ifx\csname citenamefont\endcsname\relax
  \def\citenamefont#1{#1}\fi
\expandafter\ifx\csname url\endcsname\relax
  \def\url#1{\texttt{#1}}\fi
\expandafter\ifx\csname urlprefix\endcsname\relax\def\urlprefix{URL }\fi
\providecommand{\bibinfo}[2]{#2}
\providecommand{\eprint}[2][]{\url{#2}}

\bibitem[{\citenamefont{Gong et~al.}(2017)\citenamefont{Gong, Li, Li, Ji,
  Stern, Xia, Cao, Bao, Wang, Wang et~al.}}]{Gong2017}
\bibinfo{author}{\bibfnamefont{C.}~\bibnamefont{Gong}},
  \bibinfo{author}{\bibfnamefont{L.}~\bibnamefont{Li}},
  \bibinfo{author}{\bibfnamefont{Z.}~\bibnamefont{Li}},
  \bibinfo{author}{\bibfnamefont{H.}~\bibnamefont{Ji}},
  \bibinfo{author}{\bibfnamefont{A.}~\bibnamefont{Stern}},
  \bibinfo{author}{\bibfnamefont{Y.}~\bibnamefont{Xia}},
  \bibinfo{author}{\bibfnamefont{T.}~\bibnamefont{Cao}},
  \bibinfo{author}{\bibfnamefont{W.}~\bibnamefont{Bao}},
  \bibinfo{author}{\bibfnamefont{C.}~\bibnamefont{Wang}},
  \bibinfo{author}{\bibfnamefont{Y.}~\bibnamefont{Wang}}, \bibnamefont{et~al.},
  \bibinfo{journal}{Nature} \textbf{\bibinfo{volume}{546}},
  \bibinfo{pages}{265} (\bibinfo{year}{2017}),
  \urlprefix\url{https://doi.org/10.1038/nature22060}.

\bibitem[{\citenamefont{Huang et~al.}(2017)\citenamefont{Huang, Clark,
  Navarro-Moratalla, Klein, Cheng, Seyler, Zhong, Schmidgall, McGuire, Cobden
  et~al.}}]{Huang2017}
\bibinfo{author}{\bibfnamefont{B.}~\bibnamefont{Huang}},
  \bibinfo{author}{\bibfnamefont{G.}~\bibnamefont{Clark}},
  \bibinfo{author}{\bibfnamefont{E.}~\bibnamefont{Navarro-Moratalla}},
  \bibinfo{author}{\bibfnamefont{D.~R.} \bibnamefont{Klein}},
  \bibinfo{author}{\bibfnamefont{R.}~\bibnamefont{Cheng}},
  \bibinfo{author}{\bibfnamefont{K.~L.} \bibnamefont{Seyler}},
  \bibinfo{author}{\bibfnamefont{D.}~\bibnamefont{Zhong}},
  \bibinfo{author}{\bibfnamefont{E.}~\bibnamefont{Schmidgall}},
  \bibinfo{author}{\bibfnamefont{M.~A.} \bibnamefont{McGuire}},
  \bibinfo{author}{\bibfnamefont{D.~H.} \bibnamefont{Cobden}},
  \bibnamefont{et~al.}, \bibinfo{journal}{Nature}
  \textbf{\bibinfo{volume}{546}}, \bibinfo{pages}{270} (\bibinfo{year}{2017}),
  \urlprefix\url{https://doi.org/10.1038/nature22391}.

\bibitem[{\citenamefont{Deng et~al.}(2018)\citenamefont{Deng, Yu, Song, Zhang,
  Wang, Sun, Yi, Wu, Wu, Zhu et~al.}}]{Deng2018}
\bibinfo{author}{\bibfnamefont{Y.}~\bibnamefont{Deng}},
  \bibinfo{author}{\bibfnamefont{Y.}~\bibnamefont{Yu}},
  \bibinfo{author}{\bibfnamefont{Y.}~\bibnamefont{Song}},
  \bibinfo{author}{\bibfnamefont{J.}~\bibnamefont{Zhang}},
  \bibinfo{author}{\bibfnamefont{N.~Z.} \bibnamefont{Wang}},
  \bibinfo{author}{\bibfnamefont{Z.}~\bibnamefont{Sun}},
  \bibinfo{author}{\bibfnamefont{Y.}~\bibnamefont{Yi}},
  \bibinfo{author}{\bibfnamefont{Y.~Z.} \bibnamefont{Wu}},
  \bibinfo{author}{\bibfnamefont{S.}~\bibnamefont{Wu}},
  \bibinfo{author}{\bibfnamefont{J.}~\bibnamefont{Zhu}}, \bibnamefont{et~al.},
  \bibinfo{journal}{Nature} \textbf{\bibinfo{volume}{563}}, \bibinfo{pages}{94}
  (\bibinfo{year}{2018}),
  \urlprefix\url{https://doi.org/10.1038/s41586-018-0626-9}.

\bibitem[{\citenamefont{O'Hara et~al.}(2018)\citenamefont{O'Hara, Zhu, Trout,
  Ahmed, Luo, Lee, Brenner, Rajan, Gupta, McComb et~al.}}]{Ohara2018}
\bibinfo{author}{\bibfnamefont{D.~J.} \bibnamefont{O'Hara}},
  \bibinfo{author}{\bibfnamefont{T.}~\bibnamefont{Zhu}},
  \bibinfo{author}{\bibfnamefont{A.~H.} \bibnamefont{Trout}},
  \bibinfo{author}{\bibfnamefont{A.~S.} \bibnamefont{Ahmed}},
  \bibinfo{author}{\bibfnamefont{Y.~K.} \bibnamefont{Luo}},
  \bibinfo{author}{\bibfnamefont{C.~H.} \bibnamefont{Lee}},
  \bibinfo{author}{\bibfnamefont{M.~R.} \bibnamefont{Brenner}},
  \bibinfo{author}{\bibfnamefont{S.}~\bibnamefont{Rajan}},
  \bibinfo{author}{\bibfnamefont{J.~A.} \bibnamefont{Gupta}},
  \bibinfo{author}{\bibfnamefont{D.~W.} \bibnamefont{McComb}},
  \bibnamefont{et~al.}, \bibinfo{journal}{Nano Letters}
  \textbf{\bibinfo{volume}{18}}, \bibinfo{pages}{3125} (\bibinfo{year}{2018}),
  \urlprefix\url{https://doi.org/10.1021/acs.nanolett.8b00683}.

\bibitem[{\citenamefont{Thiel et~al.}(2019)\citenamefont{Thiel, Wang, Tschudin,
  Rohner, Guti{\'e}rrez-Lezama, Ubrig, Gibertini, Giannini, Morpurgo, and
  Maletinsky}}]{Thiel2019}
\bibinfo{author}{\bibfnamefont{L.}~\bibnamefont{Thiel}},
  \bibinfo{author}{\bibfnamefont{Z.}~\bibnamefont{Wang}},
  \bibinfo{author}{\bibfnamefont{M.~A.} \bibnamefont{Tschudin}},
  \bibinfo{author}{\bibfnamefont{D.}~\bibnamefont{Rohner}},
  \bibinfo{author}{\bibfnamefont{I.}~\bibnamefont{Guti{\'e}rrez-Lezama}},
  \bibinfo{author}{\bibfnamefont{N.}~\bibnamefont{Ubrig}},
  \bibinfo{author}{\bibfnamefont{M.}~\bibnamefont{Gibertini}},
  \bibinfo{author}{\bibfnamefont{E.}~\bibnamefont{Giannini}},
  \bibinfo{author}{\bibfnamefont{A.~F.} \bibnamefont{Morpurgo}},
  \bibnamefont{and}
  \bibinfo{author}{\bibfnamefont{P.}~\bibnamefont{Maletinsky}},
  \bibinfo{journal}{Science} \textbf{\bibinfo{volume}{364}},
  \bibinfo{pages}{973} (\bibinfo{year}{2019}), ISSN \bibinfo{issn}{0036-8075},
  \urlprefix\url{https://science.sciencemag.org/content/364/6444/973}.

\bibitem[{\citenamefont{Huang et~al.}(2018)\citenamefont{Huang, Clark, Klein,
  MacNeill, Navarro-Moratalla, Seyler, Wilson, McGuire, Cobden, Xiao
  et~al.}}]{Huang2018}
\bibinfo{author}{\bibfnamefont{B.}~\bibnamefont{Huang}},
  \bibinfo{author}{\bibfnamefont{G.}~\bibnamefont{Clark}},
  \bibinfo{author}{\bibfnamefont{D.~R.} \bibnamefont{Klein}},
  \bibinfo{author}{\bibfnamefont{D.}~\bibnamefont{MacNeill}},
  \bibinfo{author}{\bibfnamefont{E.}~\bibnamefont{Navarro-Moratalla}},
  \bibinfo{author}{\bibfnamefont{K.~L.} \bibnamefont{Seyler}},
  \bibinfo{author}{\bibfnamefont{N.}~\bibnamefont{Wilson}},
  \bibinfo{author}{\bibfnamefont{M.~A.} \bibnamefont{McGuire}},
  \bibinfo{author}{\bibfnamefont{D.~H.} \bibnamefont{Cobden}},
  \bibinfo{author}{\bibfnamefont{D.}~\bibnamefont{Xiao}}, \bibnamefont{et~al.},
  \bibinfo{journal}{Nature Nanotechnology} \textbf{\bibinfo{volume}{13}},
  \bibinfo{pages}{544} (\bibinfo{year}{2018}),
  \urlprefix\url{https://doi.org/10.1038/s41565-018-0121-3}.

\bibitem[{\citenamefont{Webster and Yan}(2018)}]{Webster2018}
\bibinfo{author}{\bibfnamefont{L.}~\bibnamefont{Webster}} \bibnamefont{and}
  \bibinfo{author}{\bibfnamefont{J.-A.} \bibnamefont{Yan}},
  \bibinfo{journal}{Phys. Rev. B} \textbf{\bibinfo{volume}{98}},
  \bibinfo{pages}{144411} (\bibinfo{year}{2018}),
  \urlprefix\url{https://link.aps.org/doi/10.1103/PhysRevB.98.144411}.

\bibitem[{\citenamefont{Yazyev and Helm}(2007)}]{Yazyev2007}
\bibinfo{author}{\bibfnamefont{O.~V.} \bibnamefont{Yazyev}} \bibnamefont{and}
  \bibinfo{author}{\bibfnamefont{L.}~\bibnamefont{Helm}},
  \bibinfo{journal}{Phys. Rev. B} \textbf{\bibinfo{volume}{75}},
  \bibinfo{pages}{125408} (\bibinfo{year}{2007}),
  \urlprefix\url{https://link.aps.org/doi/10.1103/PhysRevB.75.125408}.

\bibitem[{\citenamefont{Liu and Cheng}(2007)}]{Liu2007}
\bibinfo{author}{\bibfnamefont{R.-F.} \bibnamefont{Liu}} \bibnamefont{and}
  \bibinfo{author}{\bibfnamefont{C.}~\bibnamefont{Cheng}},
  \bibinfo{journal}{Phys. Rev. B} \textbf{\bibinfo{volume}{76}},
  \bibinfo{pages}{014405} (\bibinfo{year}{2007}),
  \urlprefix\url{https://link.aps.org/doi/10.1103/PhysRevB.76.014405}.

\bibitem[{\citenamefont{Magda et~al.}(2014)\citenamefont{Magda, Jin,
  Hagym\'{a}si, Vancs\'{o}, Osv\'{a}th, Nemes-Incze, Hwang, Bir\'{o}, and
  Tapaszt\'{o}}}]{Magda2014}
\bibinfo{author}{\bibfnamefont{G.~Z.} \bibnamefont{Magda}},
  \bibinfo{author}{\bibfnamefont{X.}~\bibnamefont{Jin}},
  \bibinfo{author}{\bibfnamefont{I.}~\bibnamefont{Hagym\'{a}si}},
  \bibinfo{author}{\bibfnamefont{P.}~\bibnamefont{Vancs\'{o}}},
  \bibinfo{author}{\bibfnamefont{Z.}~\bibnamefont{Osv\'{a}th}},
  \bibinfo{author}{\bibfnamefont{P.}~\bibnamefont{Nemes-Incze}},
  \bibinfo{author}{\bibfnamefont{C.}~\bibnamefont{Hwang}},
  \bibinfo{author}{\bibfnamefont{L.~P.} \bibnamefont{Bir\'{o}}},
  \bibnamefont{and}
  \bibinfo{author}{\bibfnamefont{L.}~\bibnamefont{Tapaszt\'{o}}},
  \bibinfo{journal}{Nature} \textbf{\bibinfo{volume}{514}},
  \bibinfo{pages}{608} (\bibinfo{year}{2014}),
  \urlprefix\url{https://doi.org/10.1038/nature13831}.

\bibitem[{\citenamefont{Dev and Reinecke}(2015)}]{Dev2015}
\bibinfo{author}{\bibfnamefont{P.}~\bibnamefont{Dev}} \bibnamefont{and}
  \bibinfo{author}{\bibfnamefont{T.~L.} \bibnamefont{Reinecke}},
  \bibinfo{journal}{Phys. Rev. B} \textbf{\bibinfo{volume}{91}},
  \bibinfo{pages}{035436} (\bibinfo{year}{2015}),
  \urlprefix\url{https://link.aps.org/doi/10.1103/PhysRevB.91.035436}.

\bibitem[{\citenamefont{Dresselhaus}(1986)}]{Dresselhaus1986}
\bibinfo{author}{\bibfnamefont{M.~S.} \bibnamefont{Dresselhaus}},
  \emph{\bibinfo{title}{Intercalation in Layered Materials}}, vol.
  \bibinfo{volume}{148} of \emph{\bibinfo{series}{NATO ASI Series}}
  (\bibinfo{publisher}{Springer, Boston, MA}, \bibinfo{year}{1986}), ISBN
  \bibinfo{isbn}{978-1-4757-5558-9}.

\bibitem[{\citenamefont{Morosan et~al.}(2007)\citenamefont{Morosan, Zandbergen,
  Li, Lee, Checkelsky, Heinrich, Siegrist, Ong, and Cava}}]{Morosan2007}
\bibinfo{author}{\bibfnamefont{E.}~\bibnamefont{Morosan}},
  \bibinfo{author}{\bibfnamefont{H.~W.} \bibnamefont{Zandbergen}},
  \bibinfo{author}{\bibfnamefont{L.}~\bibnamefont{Li}},
  \bibinfo{author}{\bibfnamefont{M.}~\bibnamefont{Lee}},
  \bibinfo{author}{\bibfnamefont{J.~G.} \bibnamefont{Checkelsky}},
  \bibinfo{author}{\bibfnamefont{M.}~\bibnamefont{Heinrich}},
  \bibinfo{author}{\bibfnamefont{T.}~\bibnamefont{Siegrist}},
  \bibinfo{author}{\bibfnamefont{N.~P.} \bibnamefont{Ong}}, \bibnamefont{and}
  \bibinfo{author}{\bibfnamefont{R.~J.} \bibnamefont{Cava}},
  \bibinfo{journal}{Phys. Rev. B} \textbf{\bibinfo{volume}{75}},
  \bibinfo{pages}{104401} (\bibinfo{year}{2007}),
  \urlprefix\url{https://link.aps.org/doi/10.1103/PhysRevB.75.104401}.

\bibitem[{\citenamefont{Bointon et~al.}(2014)\citenamefont{Bointon, Khrapach,
  Yakimova, Shytov, Craciun, and Russo}}]{Bointon2014}
\bibinfo{author}{\bibfnamefont{T.~H.} \bibnamefont{Bointon}},
  \bibinfo{author}{\bibfnamefont{I.}~\bibnamefont{Khrapach}},
  \bibinfo{author}{\bibfnamefont{R.}~\bibnamefont{Yakimova}},
  \bibinfo{author}{\bibfnamefont{A.~V.} \bibnamefont{Shytov}},
  \bibinfo{author}{\bibfnamefont{M.~F.} \bibnamefont{Craciun}},
  \bibnamefont{and} \bibinfo{author}{\bibfnamefont{S.}~\bibnamefont{Russo}},
  \bibinfo{journal}{Nano Letters} \textbf{\bibinfo{volume}{14}},
  \bibinfo{pages}{1751} (\bibinfo{year}{2014}), \bibinfo{note}{pMID: 24635686},
  \urlprefix\url{https://doi.org/10.1021/nl4040779}.

\bibitem[{\citenamefont{Kumar et~al.}(2017)\citenamefont{Kumar, Skomski, and
  Pushpa}}]{Kumar2017}
\bibinfo{author}{\bibfnamefont{P.}~\bibnamefont{Kumar}},
  \bibinfo{author}{\bibfnamefont{R.}~\bibnamefont{Skomski}}, \bibnamefont{and}
  \bibinfo{author}{\bibfnamefont{R.}~\bibnamefont{Pushpa}},
  \bibinfo{journal}{ACS Omega} \textbf{\bibinfo{volume}{2}},
  \bibinfo{pages}{7985} (\bibinfo{year}{2017}),
  \urlprefix\url{https://doi.org/10.1021/acsomega.7b01164}.

\bibitem[{\citenamefont{Li et~al.}(2019)\citenamefont{Li, Chang, Xie, Cheng,
  Yang, Chen, and Chen}}]{LiKerui2019}
\bibinfo{author}{\bibfnamefont{K.}~\bibnamefont{Li}},
  \bibinfo{author}{\bibfnamefont{T.-H.} \bibnamefont{Chang}},
  \bibinfo{author}{\bibfnamefont{Q.}~\bibnamefont{Xie}},
  \bibinfo{author}{\bibfnamefont{Y.}~\bibnamefont{Cheng}},
  \bibinfo{author}{\bibfnamefont{H.}~\bibnamefont{Yang}},
  \bibinfo{author}{\bibfnamefont{J.}~\bibnamefont{Chen}}, \bibnamefont{and}
  \bibinfo{author}{\bibfnamefont{P.-Y.} \bibnamefont{Chen}},
  \bibinfo{journal}{Advanced Electronic Materials}
  \textbf{\bibinfo{volume}{5}}, \bibinfo{pages}{1900040}
  (\bibinfo{year}{2019}),
  \urlprefix\url{https://www.onlinelibrary.wiley.com/doi/abs/10.1002/aelm.201900040}.

\bibitem[{\citenamefont{Averyanov et~al.}(2018)\citenamefont{Averyanov,
  Sokolov, Tokmachev, Parfenov, Karateev, Taldenkov, and
  Storchak}}]{Averyanov2018}
\bibinfo{author}{\bibfnamefont{D.~V.} \bibnamefont{Averyanov}},
  \bibinfo{author}{\bibfnamefont{I.~S.} \bibnamefont{Sokolov}},
  \bibinfo{author}{\bibfnamefont{A.~M.} \bibnamefont{Tokmachev}},
  \bibinfo{author}{\bibfnamefont{O.~E.} \bibnamefont{Parfenov}},
  \bibinfo{author}{\bibfnamefont{I.~A.} \bibnamefont{Karateev}},
  \bibinfo{author}{\bibfnamefont{A.~N.} \bibnamefont{Taldenkov}},
  \bibnamefont{and} \bibinfo{author}{\bibfnamefont{V.~G.}
  \bibnamefont{Storchak}}, \bibinfo{journal}{ACS Applied Materials \&
  Interfaces} \textbf{\bibinfo{volume}{10}}, \bibinfo{pages}{20767}
  (\bibinfo{year}{2018}), \bibinfo{note}{pMID: 29806934},
  \urlprefix\url{https://doi.org/10.1021/acsami.8b04289}.

\bibitem[{\citenamefont{Karpiak et~al.}(2019)\citenamefont{Karpiak, Cummings,
  Zollner, Vila, Khokhriakov, Hoque, Dankert, Svedlindh, Fabian, Roche
  et~al.}}]{Karpiak2019}
\bibinfo{author}{\bibfnamefont{B.}~\bibnamefont{Karpiak}},
  \bibinfo{author}{\bibfnamefont{A.~W.} \bibnamefont{Cummings}},
  \bibinfo{author}{\bibfnamefont{K.}~\bibnamefont{Zollner}},
  \bibinfo{author}{\bibfnamefont{M.}~\bibnamefont{Vila}},
  \bibinfo{author}{\bibfnamefont{D.}~\bibnamefont{Khokhriakov}},
  \bibinfo{author}{\bibfnamefont{A.~M.} \bibnamefont{Hoque}},
  \bibinfo{author}{\bibfnamefont{A.}~\bibnamefont{Dankert}},
  \bibinfo{author}{\bibfnamefont{P.}~\bibnamefont{Svedlindh}},
  \bibinfo{author}{\bibfnamefont{J.}~\bibnamefont{Fabian}},
  \bibinfo{author}{\bibfnamefont{S.}~\bibnamefont{Roche}},
  \bibnamefont{et~al.}, \bibinfo{journal}{2D Materials}
  \textbf{\bibinfo{volume}{7}}, \bibinfo{pages}{015026} (\bibinfo{year}{2019}),
  \urlprefix\url{https://doi.org/10.1088%2F2053-1583%2Fab5915}.

\bibitem[{\citenamefont{Avsar et~al.}(2019)\citenamefont{Avsar, Ciarrocchi,
  Pizzochero, Unuchek, Yazyev, and Kis}}]{Avsar2019}
\bibinfo{author}{\bibfnamefont{A.}~\bibnamefont{Avsar}},
  \bibinfo{author}{\bibfnamefont{A.}~\bibnamefont{Ciarrocchi}},
  \bibinfo{author}{\bibfnamefont{M.}~\bibnamefont{Pizzochero}},
  \bibinfo{author}{\bibfnamefont{D.}~\bibnamefont{Unuchek}},
  \bibinfo{author}{\bibfnamefont{O.~V.} \bibnamefont{Yazyev}},
  \bibnamefont{and} \bibinfo{author}{\bibfnamefont{A.}~\bibnamefont{Kis}},
  \bibinfo{journal}{Nature Nanotechnology} \textbf{\bibinfo{volume}{14}},
  \bibinfo{pages}{674} (\bibinfo{year}{2019}), ISSN \bibinfo{issn}{17483395},
  \urlprefix\url{http://dx.doi.org/10.1038/s41565-019-0467-1}.

\bibitem[{\citenamefont{Gao et~al.}(2017)\citenamefont{Gao, Cheng, Tian, Hu,
  Zeng, Zhang, and Zhang}}]{Gao2017}
\bibinfo{author}{\bibfnamefont{J.}~\bibnamefont{Gao}},
  \bibinfo{author}{\bibfnamefont{Y.}~\bibnamefont{Cheng}},
  \bibinfo{author}{\bibfnamefont{T.}~\bibnamefont{Tian}},
  \bibinfo{author}{\bibfnamefont{X.}~\bibnamefont{Hu}},
  \bibinfo{author}{\bibfnamefont{K.}~\bibnamefont{Zeng}},
  \bibinfo{author}{\bibfnamefont{G.}~\bibnamefont{Zhang}}, \bibnamefont{and}
  \bibinfo{author}{\bibfnamefont{Y.~W.} \bibnamefont{Zhang}},
  \bibinfo{journal}{ACS Omega} \textbf{\bibinfo{volume}{2}},
  \bibinfo{pages}{8640} (\bibinfo{year}{2017}), ISSN \bibinfo{issn}{24701343},
  \urlprefix\url{https://doi.org/10.1021/acsomega.7b01619}.

\bibitem[{\citenamefont{Zulfiqar et~al.}(2016)\citenamefont{Zulfiqar, Zhao, Li,
  Nazir, and Ni}}]{Zulfiqar2016}
\bibinfo{author}{\bibfnamefont{M.}~\bibnamefont{Zulfiqar}},
  \bibinfo{author}{\bibfnamefont{Y.}~\bibnamefont{Zhao}},
  \bibinfo{author}{\bibfnamefont{G.}~\bibnamefont{Li}},
  \bibinfo{author}{\bibfnamefont{S.}~\bibnamefont{Nazir}}, \bibnamefont{and}
  \bibinfo{author}{\bibfnamefont{J.}~\bibnamefont{Ni}},
  \bibinfo{journal}{Journal of Physical Chemistry C}
  \textbf{\bibinfo{volume}{120}}, \bibinfo{pages}{25030}
  (\bibinfo{year}{2016}), ISSN \bibinfo{issn}{19327455},
  \urlprefix\url{https://doi.org/10.1021/acs.jpcc.6b06999}.

\bibitem[{\citenamefont{Zhang et~al.}(2016)\citenamefont{Zhang, Guo, Jiang,
  Tao, Song, Li, and Huang}}]{Zhang2016}
\bibinfo{author}{\bibfnamefont{W.}~\bibnamefont{Zhang}},
  \bibinfo{author}{\bibfnamefont{H.~T.} \bibnamefont{Guo}},
  \bibinfo{author}{\bibfnamefont{J.}~\bibnamefont{Jiang}},
  \bibinfo{author}{\bibfnamefont{Q.~C.} \bibnamefont{Tao}},
  \bibinfo{author}{\bibfnamefont{X.~J.} \bibnamefont{Song}},
  \bibinfo{author}{\bibfnamefont{H.}~\bibnamefont{Li}}, \bibnamefont{and}
  \bibinfo{author}{\bibfnamefont{J.}~\bibnamefont{Huang}},
  \bibinfo{journal}{Journal of Applied Physics} \textbf{\bibinfo{volume}{120}},
  \bibinfo{pages}{013904} (\bibinfo{year}{2016}), ISSN
  \bibinfo{issn}{10897550},
  \urlprefix\url{http://dx.doi.org/10.1063/1.4955468}.

\bibitem[{\citenamefont{Manchanda et~al.}(2016)\citenamefont{Manchanda, Enders,
  Sellmyer, and Skomski}}]{Manchanda2016}
\bibinfo{author}{\bibfnamefont{P.}~\bibnamefont{Manchanda}},
  \bibinfo{author}{\bibfnamefont{A.}~\bibnamefont{Enders}},
  \bibinfo{author}{\bibfnamefont{D.~J.} \bibnamefont{Sellmyer}},
  \bibnamefont{and} \bibinfo{author}{\bibfnamefont{R.}~\bibnamefont{Skomski}},
  \bibinfo{journal}{Phys. Rev. B} \textbf{\bibinfo{volume}{94}},
  \bibinfo{pages}{104426} (\bibinfo{year}{2016}),
  \urlprefix\url{https://link.aps.org/doi/10.1103/PhysRevB.94.104426}.

\bibitem[{\citenamefont{Kar et~al.}(2019)\citenamefont{Kar, Sarkar, Pal, and
  Sarkar}}]{Kar2019}
\bibinfo{author}{\bibfnamefont{M.}~\bibnamefont{Kar}},
  \bibinfo{author}{\bibfnamefont{R.}~\bibnamefont{Sarkar}},
  \bibinfo{author}{\bibfnamefont{S.}~\bibnamefont{Pal}}, \bibnamefont{and}
  \bibinfo{author}{\bibfnamefont{P.}~\bibnamefont{Sarkar}},
  \bibinfo{journal}{Journal of Physics: Condensed Matter}
  \textbf{\bibinfo{volume}{31}}, \bibinfo{pages}{145502}
  (\bibinfo{year}{2019}),
  \urlprefix\url{https://doi.org/10.1088%2F1361-648x%2Faaff40}.

\bibitem[{\citenamefont{Bonilla et~al.}(2018)\citenamefont{Bonilla, Kolekar,
  Ma, Diaz, Kalappattil, Das, Eggers, Gutierrez, Phan, and
  Batzill}}]{Bonilla2018}
\bibinfo{author}{\bibfnamefont{M.}~\bibnamefont{Bonilla}},
  \bibinfo{author}{\bibfnamefont{S.}~\bibnamefont{Kolekar}},
  \bibinfo{author}{\bibfnamefont{Y.}~\bibnamefont{Ma}},
  \bibinfo{author}{\bibfnamefont{H.~C.} \bibnamefont{Diaz}},
  \bibinfo{author}{\bibfnamefont{V.}~\bibnamefont{Kalappattil}},
  \bibinfo{author}{\bibfnamefont{R.}~\bibnamefont{Das}},
  \bibinfo{author}{\bibfnamefont{T.}~\bibnamefont{Eggers}},
  \bibinfo{author}{\bibfnamefont{H.~R.} \bibnamefont{Gutierrez}},
  \bibinfo{author}{\bibfnamefont{M.~H.} \bibnamefont{Phan}}, \bibnamefont{and}
  \bibinfo{author}{\bibfnamefont{M.}~\bibnamefont{Batzill}},
  \bibinfo{journal}{Nature Nanotechnology} \textbf{\bibinfo{volume}{13}},
  \bibinfo{pages}{289} (\bibinfo{year}{2018}), ISSN \bibinfo{issn}{17483395},
  \urlprefix\url{http://dx.doi.org/10.1038/s41565-018-0063-9}.

\bibitem[{\citenamefont{Avsar et~al.}(2020)\citenamefont{Avsar, Cheon,
  Pizzochero, Tripathi, Ciarrocchi, Yazyev, and Kis}}]{Avsar2020}
\bibinfo{author}{\bibfnamefont{A.}~\bibnamefont{Avsar}},
  \bibinfo{author}{\bibfnamefont{C.~Y.} \bibnamefont{Cheon}},
  \bibinfo{author}{\bibfnamefont{M.}~\bibnamefont{Pizzochero}},
  \bibinfo{author}{\bibfnamefont{M.}~\bibnamefont{Tripathi}},
  \bibinfo{author}{\bibfnamefont{A.}~\bibnamefont{Ciarrocchi}},
  \bibinfo{author}{\bibfnamefont{O.~V.} \bibnamefont{Yazyev}},
  \bibnamefont{and} \bibinfo{author}{\bibfnamefont{A.}~\bibnamefont{Kis}},
  \bibinfo{journal}{Nature Communications} \textbf{\bibinfo{volume}{11}},
  \bibinfo{pages}{4806} (\bibinfo{year}{2020}), ISSN \bibinfo{issn}{20411723},
  \urlprefix\url{http://dx.doi.org/10.1038/s41467-020-18521-6}.

\bibitem[{\citenamefont{Dev and Reinecke}(2014)}]{Dev2014}
\bibinfo{author}{\bibfnamefont{P.}~\bibnamefont{Dev}} \bibnamefont{and}
  \bibinfo{author}{\bibfnamefont{T.~L.} \bibnamefont{Reinecke}},
  \bibinfo{journal}{Phys. Rev. B} \textbf{\bibinfo{volume}{89}},
  \bibinfo{pages}{035404} (\bibinfo{year}{2014}),
  \urlprefix\url{https://link.aps.org/doi/10.1103/PhysRevB.89.035404}.

\bibitem[{\citenamefont{Manchanda et~al.}(2020)\citenamefont{Manchanda, Kumar,
  and Dev}}]{Manchanda2020}
\bibinfo{author}{\bibfnamefont{P.}~\bibnamefont{Manchanda}},
  \bibinfo{author}{\bibfnamefont{P.}~\bibnamefont{Kumar}}, \bibnamefont{and}
  \bibinfo{author}{\bibfnamefont{P.}~\bibnamefont{Dev}},
  \bibinfo{journal}{Phys. Rev. B} \textbf{\bibinfo{volume}{101}},
  \bibinfo{pages}{144104} (\bibinfo{year}{2020}),
  \urlprefix\url{https://link.aps.org/doi/10.1103/PhysRevB.101.144104}.

\bibitem[{\citenamefont{Ge et~al.}(2020)\citenamefont{Ge, Luo, Lin, Shi, Liu,
  Wang, Zhang, Duan, and Wang}}]{Ge2020}
\bibinfo{author}{\bibfnamefont{J.}~\bibnamefont{Ge}},
  \bibinfo{author}{\bibfnamefont{T.}~\bibnamefont{Luo}},
  \bibinfo{author}{\bibfnamefont{Z.}~\bibnamefont{Lin}},
  \bibinfo{author}{\bibfnamefont{J.}~\bibnamefont{Shi}},
  \bibinfo{author}{\bibfnamefont{Y.}~\bibnamefont{Liu}},
  \bibinfo{author}{\bibfnamefont{P.}~\bibnamefont{Wang}},
  \bibinfo{author}{\bibfnamefont{Y.}~\bibnamefont{Zhang}},
  \bibinfo{author}{\bibfnamefont{W.}~\bibnamefont{Duan}}, \bibnamefont{and}
  \bibinfo{author}{\bibfnamefont{J.}~\bibnamefont{Wang}},
  \bibinfo{journal}{Advanced Materials} \textbf{\bibinfo{volume}{2005465}},
  \bibinfo{pages}{1} (\bibinfo{year}{2020}), ISSN \bibinfo{issn}{15214095}.

\bibitem[{\citenamefont{Kresse and Furthm\"uller}(1996)}]{Kresse1996}
\bibinfo{author}{\bibfnamefont{G.}~\bibnamefont{Kresse}} \bibnamefont{and}
  \bibinfo{author}{\bibfnamefont{J.}~\bibnamefont{Furthm\"uller}},
  \bibinfo{journal}{Phys. Rev. B} \textbf{\bibinfo{volume}{54}},
  \bibinfo{pages}{11169} (\bibinfo{year}{1996}),
  \urlprefix\url{https://link.aps.org/doi/10.1103/PhysRevB.54.11169}.

\bibitem[{\citenamefont{Giannozzi et~al.}(2009)\citenamefont{Giannozzi, Baroni,
  Bonini, Calandra, Car, Cavazzoni, Ceresoli, Chiarotti, Cococcioni, Dabo
  et~al.}}]{Giannozzi2009}
\bibinfo{author}{\bibfnamefont{P.}~\bibnamefont{Giannozzi}},
  \bibinfo{author}{\bibfnamefont{S.}~\bibnamefont{Baroni}},
  \bibinfo{author}{\bibfnamefont{N.}~\bibnamefont{Bonini}},
  \bibinfo{author}{\bibfnamefont{M.}~\bibnamefont{Calandra}},
  \bibinfo{author}{\bibfnamefont{R.}~\bibnamefont{Car}},
  \bibinfo{author}{\bibfnamefont{C.}~\bibnamefont{Cavazzoni}},
  \bibinfo{author}{\bibfnamefont{D.}~\bibnamefont{Ceresoli}},
  \bibinfo{author}{\bibfnamefont{G.~L.} \bibnamefont{Chiarotti}},
  \bibinfo{author}{\bibfnamefont{M.}~\bibnamefont{Cococcioni}},
  \bibinfo{author}{\bibfnamefont{I.}~\bibnamefont{Dabo}}, \bibnamefont{et~al.},
  \bibinfo{journal}{Journal of Physics: Condensed Matter}
  \textbf{\bibinfo{volume}{21}}, \bibinfo{pages}{395502}
  (\bibinfo{year}{2009}),
  \urlprefix\url{https://doi.org/10.1088/0953-8984/21/39/395502}.

\bibitem[{\citenamefont{Perdew et~al.}(1996)\citenamefont{Perdew, Burke, and
  Ernzerhof}}]{Perdew1996}
\bibinfo{author}{\bibfnamefont{J.~P.} \bibnamefont{Perdew}},
  \bibinfo{author}{\bibfnamefont{K.}~\bibnamefont{Burke}}, \bibnamefont{and}
  \bibinfo{author}{\bibfnamefont{M.}~\bibnamefont{Ernzerhof}},
  \bibinfo{journal}{Phys. Rev. Lett.} \textbf{\bibinfo{volume}{77}},
  \bibinfo{pages}{3865} (\bibinfo{year}{1996}),
  \urlprefix\url{https://link.aps.org/doi/10.1103/PhysRevLett.77.3865}.

\bibitem[{\citenamefont{Grimme et~al.}(2010)\citenamefont{Grimme, Antony,
  Ehrlich, and Krieg}}]{Grimme2010}
\bibinfo{author}{\bibfnamefont{S.}~\bibnamefont{Grimme}},
  \bibinfo{author}{\bibfnamefont{J.}~\bibnamefont{Antony}},
  \bibinfo{author}{\bibfnamefont{S.}~\bibnamefont{Ehrlich}}, \bibnamefont{and}
  \bibinfo{author}{\bibfnamefont{H.}~\bibnamefont{Krieg}},
  \bibinfo{journal}{The Journal of Chemical Physics}
  \textbf{\bibinfo{volume}{132}}, \bibinfo{pages}{154104}
  (\bibinfo{year}{2010}), \urlprefix\url{https://doi.org/10.1063/1.3382344}.

\bibitem[{\citenamefont{Ciarrocchi et~al.}(2018)\citenamefont{Ciarrocchi,
  Avsar, Ovchinnikov, and Kis}}]{Ciarrocchi2018}
\bibinfo{author}{\bibfnamefont{A.}~\bibnamefont{Ciarrocchi}},
  \bibinfo{author}{\bibfnamefont{A.}~\bibnamefont{Avsar}},
  \bibinfo{author}{\bibfnamefont{D.}~\bibnamefont{Ovchinnikov}},
  \bibnamefont{and} \bibinfo{author}{\bibfnamefont{A.}~\bibnamefont{Kis}},
  \bibinfo{journal}{Nature Communications} \textbf{\bibinfo{volume}{9}},
  \bibinfo{pages}{1} (\bibinfo{year}{2018}), ISSN \bibinfo{issn}{20411723},
  \urlprefix\url{https://doi.org/10.1038/s41467-018-03436-0}.

\bibitem[{\citenamefont{Villaos et~al.}(2019)\citenamefont{Villaos, Crisostomo,
  Huang, Huang, Padama, Albao, Lin, and Chuang}}]{Villaos2019}
\bibinfo{author}{\bibfnamefont{R.~A.~B.} \bibnamefont{Villaos}},
  \bibinfo{author}{\bibfnamefont{C.~P.} \bibnamefont{Crisostomo}},
  \bibinfo{author}{\bibfnamefont{Z.~Q.} \bibnamefont{Huang}},
  \bibinfo{author}{\bibfnamefont{S.~M.} \bibnamefont{Huang}},
  \bibinfo{author}{\bibfnamefont{A.~A.~B.} \bibnamefont{Padama}},
  \bibinfo{author}{\bibfnamefont{M.~A.} \bibnamefont{Albao}},
  \bibinfo{author}{\bibfnamefont{H.}~\bibnamefont{Lin}}, \bibnamefont{and}
  \bibinfo{author}{\bibfnamefont{F.~C.} \bibnamefont{Chuang}},
  \bibinfo{journal}{npj 2D Materials and Applications}
  \textbf{\bibinfo{volume}{3}}, \bibinfo{pages}{1} (\bibinfo{year}{2019}),
  \urlprefix\url{http://dx.doi.org/10.1038/s41699-018-0085-z}.

\bibitem[{\citenamefont{Zhang et~al.}(2017)\citenamefont{Zhang, Qin, Huang, and
  Zhang}}]{Zhang2017}
\bibinfo{author}{\bibfnamefont{W.}~\bibnamefont{Zhang}},
  \bibinfo{author}{\bibfnamefont{J.}~\bibnamefont{Qin}},
  \bibinfo{author}{\bibfnamefont{Z.}~\bibnamefont{Huang}}, \bibnamefont{and}
  \bibinfo{author}{\bibfnamefont{W.}~\bibnamefont{Zhang}},
  \bibinfo{journal}{Journal of Applied Physics} \textbf{\bibinfo{volume}{122}},
  \bibinfo{pages}{205701} (\bibinfo{year}{2017}),
  \urlprefix\url{https://doi.org/10.1063/1.5000419}.

\bibitem[{\citenamefont{Henkelman et~al.}(2006)\citenamefont{Henkelman,
  Arnaldsson, and J{\'{o}}nsson}}]{Henkelman2006}
\bibinfo{author}{\bibfnamefont{G.}~\bibnamefont{Henkelman}},
  \bibinfo{author}{\bibfnamefont{A.}~\bibnamefont{Arnaldsson}},
  \bibnamefont{and}
  \bibinfo{author}{\bibfnamefont{H.}~\bibnamefont{J{\'{o}}nsson}},
  \bibinfo{journal}{Computational Materials Science}
  \textbf{\bibinfo{volume}{36}}, \bibinfo{pages}{354} (\bibinfo{year}{2006}),
  ISSN \bibinfo{issn}{09270256},
  \urlprefix\url{http://www.sciencedirect.com/science/article/pii/S0927025605001849}.

\bibitem[{\citenamefont{Dev et~al.}(2008)\citenamefont{Dev, Xue, and
  Zhang}}]{Dev2008}
\bibinfo{author}{\bibfnamefont{P.}~\bibnamefont{Dev}},
  \bibinfo{author}{\bibfnamefont{Y.}~\bibnamefont{Xue}}, \bibnamefont{and}
  \bibinfo{author}{\bibfnamefont{P.}~\bibnamefont{Zhang}},
  \bibinfo{journal}{Phys. Rev. Lett.} \textbf{\bibinfo{volume}{100}},
  \bibinfo{pages}{117204} (\bibinfo{year}{2008}),
  \urlprefix\url{https://link.aps.org/doi/10.1103/PhysRevLett.100.117204}.

\bibitem[{\citenamefont{Dev and Zhang}(2010)}]{Dev2010}
\bibinfo{author}{\bibfnamefont{P.}~\bibnamefont{Dev}} \bibnamefont{and}
  \bibinfo{author}{\bibfnamefont{P.}~\bibnamefont{Zhang}},
  \bibinfo{journal}{Phys. Rev. B} \textbf{\bibinfo{volume}{81}},
  \bibinfo{pages}{085207} (\bibinfo{year}{2010}),
  \urlprefix\url{https://link.aps.org/doi/10.1103/PhysRevB.81.085207}.

\bibitem[{\citenamefont{Dev}(2020)}]{Dev2020}
\bibinfo{author}{\bibfnamefont{P.}~\bibnamefont{Dev}}, \bibinfo{journal}{Phys.
  Rev. Research} \textbf{\bibinfo{volume}{2}}, \bibinfo{pages}{022050}
  (\bibinfo{year}{2020}),
  \urlprefix\url{https://link.aps.org/doi/10.1103/PhysRevResearch.2.022050}.

\bibitem[{\citenamefont{Kumar et~al.}(2020)\citenamefont{Kumar, Naumov,
  Manchanda, and Dev}}]{kumar2020}
\bibinfo{author}{\bibfnamefont{P.}~\bibnamefont{Kumar}},
  \bibinfo{author}{\bibfnamefont{I.~I.} \bibnamefont{Naumov}},
  \bibinfo{author}{\bibfnamefont{P.}~\bibnamefont{Manchanda}},
  \bibnamefont{and} \bibinfo{author}{\bibfnamefont{P.}~\bibnamefont{Dev}},
  \emph{\bibinfo{title}{A new class of intrinsic magnet: two-dimensional
  yttrium sulphur selenide}} (\bibinfo{year}{2020}), \eprint{arXiv:2009.07232}.

\end{thebibliography}

\end{document}